# Mining Essential Relationships under Multiplex Networks

Liu Weiyi, Hu Guangmin
School of Communication and Information Engineering
University of Electronic Science and Technology of China (UESTC)
Chengdu, Sichuan, China
unique_liu@163.com, hgm@uestc.edu.cn

Chen Lingli
School of Management and Economics of UESTC
University of Electronic Science and Technology of China (UESTC)
Chengdu, Sichuan, China
lingli324@163.com

*Abstract*—In big data times, massive datasets often carry different relationships among the same group of nodes, analyzing on these heterogeneous relationships may give us a window to peek the essential relationships among nodes. In this paper, first of all we propose a new metric "similarity rate" in order to capture the changing rate of similarities between node-pairs though all networks; secondly, we try to use this new metric to uncover essential relationships between node-pairs which essential relationships are often hidden and hard to get. From experiments study of Indonesian Terrorists dataset, this new metric similarity rate function well for giving us a way to uncover essential relationships from lots of appearances.

## I. INTRODUCTION

With the growing popularity of the Internet and the advent of the era of big data, massive datasets can exhibit relationships diversity when observing same group of nodes (e.g. a bunch of people from Facebook). The reason why such circumstance occurs is mainly because these datasets represent relationships from different angles, and such diversity is just reflections of the essential relationship among these nodes. Mining on them may give us a window to peek the relationship essentiality among nodes. Take online social network as an example, it simply contains three basic relationships among nodes: "retweet relationships", "comment relationships" and "attention relationships", where "retweet relationships" is mainly because of two friends' interaction behavior; "comment relationships" is mainly because two friends share same interests and "attention relationships" is mainly because of the recognition among them. In other words, the essential relationship is friendship, where "attention", "retweet" and "comment" are just reflections of such essentiality.

Much works have been done to uncover the essential relationships between nodes, one of many key methods is group detection, which try to group nodes based on similar interests. Because of big data time arriving, such methods pay much more attention concentrating on multiplex networks (Where two nodes are connected by more than one connection, relation, tie)[1]. In 2009, Tang[3] proposed Feature integration to abstract every structural feature for each network, then use them to do feature integration to form a new network. In 2010, Mucha[4] regarded single network as slice, and used inter-slice weight $s_{ip}$, cross-slice strength $c_{ip}$ and total strength $w_{ip} = s_{ip} + c_{ip}$ to depict different relationships between these slice, then generalized modularity[8][9][10][11] into multiplex networks. In 2011, based on their work, Carchiolo[2] applied Arenas's theory of "network merging will cause no change for modularity"[12][13] to multiplex networks, firstly they prove Arenas's theory remain unchanged in multiplex networks, then they generalized BGLL[14] into multiplex networks to do group detection. Magnani use multi-layer distances to do network integration in 2013 [15][16], and in 2014, Guang[17] proposed inter simplex to deal with network importance $I_s$ among multiplex networks, and use similarity weight $A_{(u,v)}^{(i)}$ to depict node-pair $(u,v)$'s similarity weight in $i^{th}$ network, then based on these two measurements, they gave a unified adjacent matrix $L = \sum_{i}^{m} I_i A^{(i)}$ to depict aggregated network, then using group detection to find essential relationships among nodes

As is known to all, relations diversity plays a majority role in forming multiplex networks, but through all these literatures, it's easily to tell that the main idea of group detection in multiplex networks is to do edges (relations) aggregation, which may certainly lose some important information about the essence why two nodes have multiple relationships. Although these works achieve a lot in finding essential relationships, but they don't take the basic knowledge that edges in multiplex networks may vary different one another, in other words, lack of considering edge heterogeneity may become the bottleneck of the existing researches. Aiming to solve this problem, in this paper, we propose a new metric "similarity rate" to take heterogeneous edges into consideration, and using such metric to try to mine essential relationships among nodes.

Section II we give a brief overview of what is "similarity rate" and why this metric can preserve relationships diversity when analyzing multiplex networks, and we describe how to use this new metric to detect groups in multiplex networks in section III. Section IV introduces datasets and represents results about how similarity rate functions in group detection. Section V concludes the paper.

## II. SIMILARITY RATE METRIC

From complex network analysis, node-pair similarity is a major weapon in depicting nodes pair relationships quantitatively, and local similarity metric is widely accepted and used in many domains because of its feasibility and applicability [18]. In this section, we begin by showing the reason why traditional similarity metric fail for describing node-pair similarities in multiplex networks, then introduce how similarity rate metric can depict node-pair similarity in multiplex networks.

### A. Traditional similarity metric failure

. In simplex network, traditional similarity metric such as *Jaccard Similarity Metric* is widely used to depict node similarity because there is only one network two nodes may involve in, which means there is only one number to describe two nodes' local similarity. But situation is quite different in multiplex network: a same group of nodes may involve in multiple networks, which means there may remain multiple local similarities between them. From section I we can tell that the mainstream of the existing methods tend to aggregate all these nodes' neighbors together, and it's quite simple to deduce that more neighbors two nodes may share, the larger local similarity may become. This is quite similar to the very nature of velocity: faster the speed is, better the performance is. So we think the physical backgrounds of local similarity and velocity remains the same.

Take Fig. 1 as an example, two cars (a classic one and a sports one) stand for two nodes in a network, time of them to through the finish line $t_1$ and $t_2$ can be used to stand for numbers of networks two nodes may involve in, and distance $s$ can be used to depict local similarity. In order to discuss the performance of these cars, we may use velocity $v$ instead of distance $s$, and it is quite obvious to find $t_1 > t_2$ for $v_1 < v_2$. This is analogue to calculate nodes' similarity in network(s). For simplex network, because there is only one network two nodes may involve in, which means time $t$ equals to 1, and distance $s$ is identically equal to local similarity $v$ for $s = v$. But for multiplex networks, time $t$ equals to the number of multiplex networks, which lead to $s = v \cdot t$, that means if we still use distance $s$ to stand for local similarity, there may occur a mistake for $v_1 < v_2$ but $s_1 = s_2$ for $t_1 > t_2$ as shown in the fig.

From network analysis perspective, the reason why local similarity may success in simplex network is because there is only one network two nodes may involve in, and the edge remains homogeneously in such network, but when considering multiplex networks, this mechanism is quite poor because it lacks of considering edge heterogeneity in different networks.

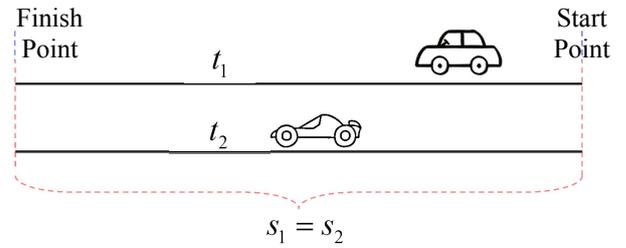

Fig. 1. Analogy example of car performance and local similarity

### B. Similarity rate metric

The weakness of traditional similarity metric push us to look back into circumstance showed in Fig. 1, and it's quite easy to find that although these two cars share the same distance, but the rate of change of distance, a.k.a Velocity, can be used to compare two cars performance. In node-pair similarity case, we think it's nature to use the rate of change of local similarities in multiple networks, as we call it "similarity rate". Because there are multiple local similarities when a node involves in multiple networks, the goal of similarity rate metric is to measure the changing rate of these local similarities. If such rate weren't change too much in all these multiple networks, it'll reveal the fact that such node-pair's relationship is solid and persistent, which may indicate that these two nodes may share the same essential relationship in multiplex networks as they may have same interests. In conclusion, similarity rate metric is proposed to quantity the changing rate of relationships between nodes: more rapid changing of local similarity is, the larger the similarity rate becomes which lead to large possibility that two nodes may share the same essentiality, vice versa.

Because velocity can be seen as the slope of distance and time in (1), it's convenience for us to do analogue on similarity rate. Similarity rate $r(u,v)$ of node-pair $(u,v)$ can also be seen as the slope of local similarities $\vec{s}(u,v)$ in all networks and the number of networks $|\vec{G}|$ in multiplex networks, as shown in (2).

$$v = {ds(t)}/{dt} \qquad (1)$$

$$r(u,v) = {ds(u,v)}/{d|\vec{G}|} \qquad (2)$$

Take Fig. 2 as an example, the number of networks equals to four, horizontal axis stands for the number of networks, vertical axis stands for local similarity of node-pair $(u,v)$ in (a) and $(u',v')$ in (b), and four circles stand for four local similarities in each network. We use Least Squares fitting to find the linear function $y = kx + b$ and $y = k'x + b'$ to depict these local similarities, shown as solid lines upon circles, where the slope of such linear function may indicate the similarity rate between node-pair. From this fig, it is quiet easy to tell that local similarities change mildly in (a) while changing dramatically in (b), which may suggest that the less possibility that node-pair $(u',v')$ may share the same relationship than node-pair $(u,v)$.

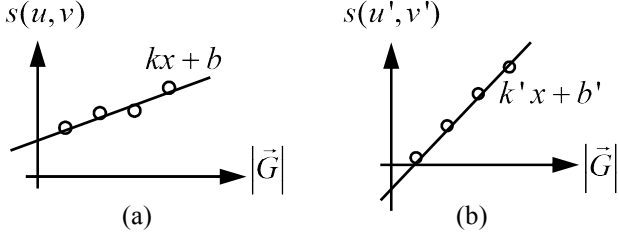

Fig. 2. Toy example of how slope infect similarity rate

There may occur a problem when only using slope to stands for similarity rate between node-pairs, if two slops remained same, there is no chance in distinguish them. In order to solve this weakness, we take the influence of intercept $b$ in the consideration at the same time. It's obviously to find that unlike slope, there may remain two situations the intercept may have: $b>0$ and $b<0$. Take Fig. 3 as an example, the legends remain the same with Fig. 2. It's quiet clearly to show that intercepts can also infect similarity rate on the condition where slopes remain the same. Because for $b>0$ or $b<0$ as shown in sub figs, the larger the intercept may be, the more possibility that node-pair $(u,v_1)$ may share the same community than node-pair $(u,v_2)$, as larger intercept may have larger local similarities, vice versa.

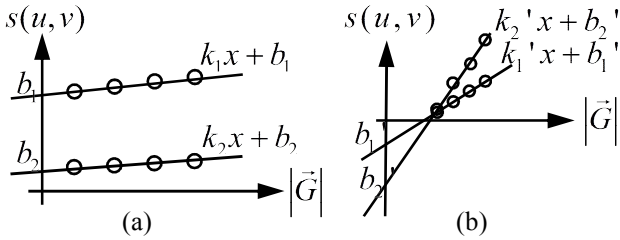

Fig. 3. Toy example of how intercept infect similarity rate

From the above, it's clear to tell that similarity rate is in direct proportion to intercept and is inversely proportional to slope. Here, we use equation (3) to give the definition of node-pair $(u,v)$ similarity rate $r(u,v)$ on multiplex networks, in order to avoid any negative number, we reassign similarity rate in (4).

$$r(u,v) \propto b/k \qquad (3)$$

$$r(u,v) = \exp(b/k) \qquad (4)$$

## III. GROUP DETECTION USING SIMILARITY RATE

For group detection on multiplex networks, the most import thing of it is to form a new network, and to do group detection on this new network [19]. In here, similarity rate can be naturally used in network integration because we can form a weighted network based on multiple networks where weight between node-pairs refers to similarity rates. After obtaining the weighted network, lots of group detection methods such as weighted modularity can be applied to find groups in multiplex networks. In below, we briefly give main steps of how to use similarity rate to do group detection upon multiplex networks.

Give Multiplex Networks $MN = (V,E,L)$, where $V$ stands for nodes involving in all networks; $L$ stands for each network under multiplex networks; $E = <x,y,l>$ stands for node-pair $(x,y)$ in network $l$.

For each node-pair $(x,y) \in V \times V$: first of all, Calculate local similarities $\overrightarrow{LS(x,y)}$ as local similarity across all networks in multiplex networks, as shown in (5), where $ls(x,y)^{l_i}$ stands for local similarity of node-pair $(x,y)$ in network $l_{i,i \in L}$; secondly, ranging $\overrightarrow{LS(x,y)}$ ascendingly to make sure $k \geq 0$ and using Least Squares fitting to find the linear function $y(x,y) = k_{(x,y)}x + b_{(x,y)}$; thirdly, using equation (4) to calculate similarity rate of node-pair $(x,y)$, clearly there remains four cases of slope and intercept.

- $k_{(x,y)} = 0, b_{(x,y)} = 0$: $r(x,y) = 0$;
- $k_{(x,y)} > 0, b_{(x,y)} > 0$: $r(x,y) = \exp(b/k)$;
- $k_{(x,y)} > 0, b_{(x,y)} = 0$: $r(x,y) = \exp(1/k)$;
- $k_{(x,y)} > 0, b_{(x,y)} < 0$: $r(x,y) = \exp(b/k)$

After finding all node-pairs' similarity rate among multiplex networks, we do network integration to form a new network $G = (V,E')$ where $V$ remains the same in $MN$, and the edge's weight $e(x,y) \in E$ in this new network can be assigned in (6).

$$\overrightarrow{LS(x,y)} = [ls(x,y)^{l_1},...,ls(x,y)^{l_i},...,ls(x,y)^{l_{|L|}}] \quad (5)$$

$$e(x,y) = r(x,y) \qquad (6)$$

Finally, lots of group detection methods such as weighted modularity can be used in finding groups upon this weighted network $G = (V,E')$.

## IV. EXPERIMENT STUDY & RESULTS

In this section, we mainly focus on describing the dataset we collected for experiments and results of mining essential relationships among them.

### A. Data sets Introduction

In order to have some ordinary relationships between nodes and have some essential relationships such as kinship relationships, a famous multiplex network data set, "Indonesian terrorist" [20] comes into our mind. It has 78 Indonesian terrorists where constituting 13 networks. From TABLE I. we can easily tell that network 11~13 maybe the essential relationships between these terrorists, and networks 1~10 maybe the reflection of such essentiality.

TABLE I.  INDONESIAN TERRORIST MULTIPLEX NETWORKS

| | Networks | Nodes number | Edges number |
|---|---|---|---|
| 1. | business | 13 | 15 |
| 2. | classmates | 39 | 175 |
| 3. | education | 37 | 284 |
| 4. | logistical | 31 | 82 |
| 5. | meeting | 26 | 63 |
| 6. | operations | 40 | 267 |
| 7. | organization | 64 | 416 |
| 8. | religious | 12 | 12 |
| 9. | training | 39 | 147 |
| 10. | communication | 75 | 201 |
| 11. | friendship | 62 | 93 |
| 12. | kinship | 24 | 16 |
| 13. | soulmates | 9 | 11 |

### B. Processing

Unlike most processing methods treating these 13 networks into four different multiplex networks, such as T (Trust), B (Business), O (Operation) and C (Communication), we divide these networks into two sets, aggregated reflection network and essentiality networks. In below, firstly give details of how to deal with these two datasets, then give an explanation on why choose community as the validation feature for uncovering the essential relationships or not.

1). Aggregated Reflection Network: because network 1~10 can all been seen as the reflections of networks 11~13, we use similarity rate discussed in section III to find all node-pairs' similarity rate among these networks, then do network integration to form a new weighted network as Aggregated Reflection Network.

2). Essentiality Network: it can be generated as aggregating network 11~13 together, where weight "1" for edges in friendship network, weight "2" for edges in kinship network and weight "3.5" for edges in soulmates network. This means, if a node-pair appear both in friendship network and soulmates network, then the weight of such node-pair will be $5.5=(2+3.5)$ in essentiality network. Using such mechanism, we can get an essentiality network with 68 nodes and 117 edges, Fig. 4 demonstrates such network. It's clearly to find there may remain multiple communities in it.

With the purpose to uncover essential relationships among nodes, we take Aggregated Reflection Network as experiment network, and Essentiality Network as ground truth network. In other words, we try to use similarity rate from reflection networks to reveal some solid pattern in essentiality network. As discussed before, communities represent a group of nodes who share same interests. So we treat communities as the solid pattern the essentiality network may reveal, and use weighted modularity to find communities in aggregated reflection network and essentiality network, and try to compare if communities in essentiality network may find themselves in aggregated reflection network.

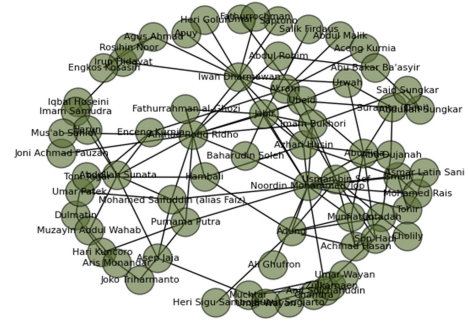

Fig. 4. Essentiality Network aggregated by network 11~13

### C. Validation

In order to do validations, using $EX_{com}$ for Experiment Community that stands for communities from experiment network, and $ES_{com}$ for Essentiality Community that stands for communities from essentiality network, then we propose overlapping rate $O$ on communities in (6) for quantify how many nodes are found in Essentiality communities while they are in the same community in experiment network. Obviously, the larger overlapping rate is, the closer two community may become.

$$O = \frac{|EX_{com} \cap ES_{com}|}{|ES_{com}|} \quad (6)$$

Fig.5 shows the validation result, horizontal axes contains 14 communities in aggregated reflection network, vertical axes means the overlapping rate for each community. Highest overlapping rate is 0.979 for community 1, which means a solid pattern of these nodes can be found which belongs to community 1; on the contrary, the lowest overlapping rate is 0.167 for community 13, which means it's failed to uncover the solid pattern in such community. In average, overlapping rate for all these communities is 0.387. Next, we try to evaluate if such rate might uncover some patterns in essentiality network.

In Fig.6, horizontal axes stands for all 116 edges from essentiality network, and vertical axes gives the frequency if such edge may find in network 1~10. The average frequency equals to 2.838, which is quite smaller than the number of multiplex network (network 1~10). It means for network 1~10, they do not contain much information about essential relationships among nodes, such phenomenon may contain a problem when mining solid patterns in essential network, because edges from multiplex network may not carry as much necessary information as to uncover the essential relationships. For example in online social network, if there were no edges for two friends in "attention", "retweet" and "comment" networks, there is too hard to judge these two nodes are friends.

In conclusion, the averaging overlapping rate 0.387 is quite good for convincing us to believe similarity rate metric may actually uncover some essential relationships among nodes.

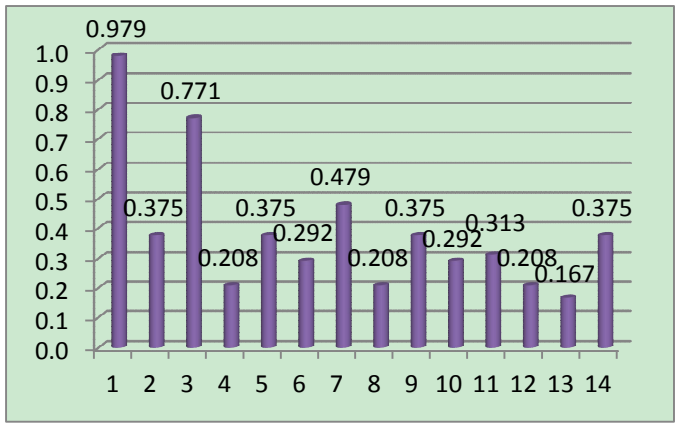

Fig.5. Validation Result

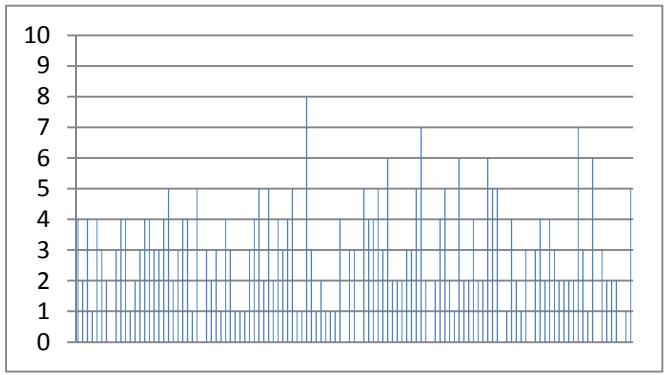

Fig.6. Per Edge Finding in network 1~10

## V. CONCLUTION

In this paper, we propose a new metric "similarity rate" for preserving edge differences in multiplex networks, which original local similarity may fail for depicting node-pair relationships. This new metric mainly concentrate on capturing the changing rate of node-pair local similarities though all networks in multiplex network, which is quit analogue to the idea of velocity. Then we use this new metric to find out if it could uncover some essential relationships among nodes. At last, we use Indonesia Terrorists Datasets to demonstrate the validity of similarity rate that this metric may indeed reveal some essential relationships in multiplex network.

In the future, we think more datasets must be used to discuss the effectiveness of similarity rate, and also have to combine such metric with some machine learning method for mining larger and more accurate essential relationships among nodes.

ACKNOWLEDGMENT

Thanks for all advices coming from anonymous referees, and this work was supported by National Natural Science Foundation of China (No.61301274).